\documentclass[11pt]{article}
\font\sixrm=cmr6
\usepackage{sint,macros}
\usepackage{amssymb}
\usepackage[dvips]{epsfig}
\begin{document}
\begin{titlepage}

\begin{flushright}
MS-TP-01-1\\
CERN-TH/2001-224\\
Bicocca-FT-01-13
\end{flushright}

\vskip 1 cm
\begin{center}
{\Large\bf 
Cutoff effects in twisted mass lattice QCD
}
\end{center}
\vskip 1 cm
\vbox{
\centerline{
\epsfxsize=2.5 true cm
\epsfbox{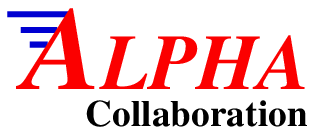}}
}
\vskip 1 cm
\begin{center}
{\large
Michele Della Morte$^{\scriptscriptstyle a}$, 
Roberto Frezzotti$^{\scriptscriptstyle a,c}$,\\
Jochen Heitger$^{\scriptscriptstyle b}$ and
Stefan Sint$^{\scriptscriptstyle c}$
}
\vskip 1.5cm
$^{\scriptstyle a}$
Universit\`a di Milano-Bicocca, Dipartimento di Fisica,\\
Piazza della Scienza 3, I-20126 Milano, Italy
\vskip 2.5ex
$^{\scriptstyle b}$
Westf\"alische Wilhelms-Universit\"at M\"unster,
Institut f\"ur Theoretische Physik,\\
Wilhelm-Klemm-Strasse 9, D-48149 M\"unster, Germany
\vskip 2.5ex
$^{\scriptstyle c}$
CERN, Theory Division,\\
CH-1211 Geneva 23, Switzerland
\vskip 1.5cm
{\bf Abstract}
\vskip 0.7ex
\end{center}

We present a first numerical study of lattice QCD with 
O($a$) improved Wilson quarks and a chirally twisted mass term.
Renormalized correlation functions are derived from the Schr\"odinger
functional and evaluated  in an intermediate space-time volume of size
$0.75^3\times 1.5 \; {\rm fm}^4$. In the quenched approximation precise
results are then obtained with a moderate computational effort,
allowing for a detailed study of the continuum approach. The latter is
discussed in terms of observables which converge to meson masses and decay
constants in the limit of large space-time volume. In the O($a$) improved
theory we find residual cutoff effects to be at the level of a few percents
at $a\simeq 0.1\,{\rm fm}$.


\vskip 1.5ex
\vfill

\begin{center}
August 2001
\end{center}

\eject
\vfill
\eject
\end{titlepage}

\section{Introduction}
Lattice twisted mass QCD (lattice tmQCD) has been introduced in 
Refs.~\cite{FGSW,paper1} as a solution to the problem of spurious
quark zero modes, which plague lattice computations with light quarks 
of the Wilson type, especially if the action is ${\rm O}(a)$ improved.
The occurrence of spurious quark zero modes causes a breakdown of the 
quenched and partially quenched approximations, as well as technical 
problems in the fully unquenched simulations. In Ref.~\cite{paper2}
it has been shown how Symanzik's on-shell improvement programme \cite{Sym} can
be implemented in the framework of this new lattice regularization,
which is intended for QCD with two mass degenerate quarks. 

The purpose of this paper is to investigate the scaling violations in
lattice twisted mass QCD, following the lines of the study presented in
Ref.~\cite{Jo} for the ${\rm O}(a)$ improved Wilson lattice action.
The tmQCD lattice regularization differs from the latter only by the
parameterization of the quark mass term, which is rotated in the chiral
flavour space, whereas the Wilson term remains in the standard form. 
This difference is the key to
avoid spurious quark zero modes and can viewed as a change of the quark field
basis that leaves unchanged --up to cutoff effects-- the physical content
of the theory. Therefore it often happens that a certain physical quantity is
obtained from a different correlation function --with different cutoff effects--
compared to the case of standard quark mass parameterization.
Moreover all the cutoff effects that
are proportional to (some power of) the quark mass may quantitatively change,
although for light quarks this is expected to be a small effect.
These remarks motivate our investigation of scaling violations in lattice tmQCD. 
 
The physical parameters of the present study have been chosen so as to be in
a situation similar to that of the scaling test of Ref.~\cite{Jo}.
Namely, we consider a system of finite size, $(L^3 \times T \simeq 0.75^3 \times 1.5)
\,{\rm fm}^4$, with Schr\"odinger functional boundary conditions, and give the
relevant renormalized quark mass parameter, $M_\rmR$, a value such that $LM_\rmR \sim 0.15$.
For such a system, we study the approach to the continuum limit of a few renormalized
observables, which in the limit $T \to \infty$ have the same physical interpretation
as the observables studied in Ref.~\cite{Jo}. When also the limit of large $L$ is taken, 
the observables turn into the pion mass, the $\rho$-meson mass, the pion decay constant
and a quantity related to the $\rho$-meson decay constant. Based on the study of
Ref.~\cite{Garden_et_al} in large spatial volume, we expect the cutoff effects observed 
at $L=0.75 \,{\rm fm}$ to be indicative of the size of the lattice artifacts in infinite volume.

In Section 2 we introduce the relevant correlation functions within the
Schr\"odinger functional setup for lattice tmQCD. The renormalized observables
of interest are constructed in Section 3, where the renormalization scheme
adopted for tmQCD is also specified. Section 4 presents numerical details
and a discussion of our results, while conclusions are drawn in Section 5.  
A preliminary report on the present work has already appeared in Ref.~\cite{proc00}.
In the following we assume that the reader is familiar with 
Refs.~\cite{FGSW,paper1,paper2}
and refer to the equations of Ref.~\cite{paper2} by using the prefix I.

\section{Schr\"odinger functional correlation functions}

The Schr\"odinger functional (SF) for lattice tmQCD has been introduced
in Ref.~\cite{paper2}, where it is defined as the integral kernel
of the integer power $T/a$ of the transfer matrix. It admits the
following representation:
\begin{equation} \label{SF_LtmQCD}
  {\cal Z}[\rhoprime,\rhobarprime,C'; \rho,\rhobar,C] =
   \int \rmD[U]\rmD[\psi] \rmD[\psibar]\,\, \rme^{-S[U,\psibar,\psi]},
\end{equation}
where $S[U,\psibar,\psi]$ is the Euclidean action of tmQCD and the
arguments of ${\cal Z}$ are the prescribed boundary values of the
gauge and quark fields at $x_0 = 0$ ($C$, $\rho$, $\rhobar$) and 
$x_0 = T$ ($C'$, $\rhoprime$, $\rhobarprime$).
As usual, the Dirichlet time-boundary conditions for the quark fields
involve the projectors $P_\pm = (1 \pm \gamma_0)/2$.

Renormalizability and ${\rm O}(a)$ improvement of the SF for tmQCD
have been discussed in Section 3 of Ref.~\cite{paper2}. We take over
the outcome of that discussion and extend the action $S[U,\psibar,\psi]$
to include all the counterterms that are needed for renormalization and 
${\rm O}(a)$ improvement, as detailed in eq.~(I.3.4).
In particular, adopting the same notational conventions 
as in Refs.~\cite{paper2,O(a)}, the quark action $S_{\rm F}[U,\psibar,\psi]$
takes the same form as on the infinite lattice:
\begin{equation}
  S_{\rm F}[U,\psibar,\psi] = a^4\sum_{x}
  \psibar(x)\left(D+\delta D+m_0+i\muq\gamma_5\tau^3\right)\psi(x) ,
  \label{tmLQCD}
\end{equation}
where $\delta D$ stands for the sum of the volume and the boundary
O($a$) counterterms.

The SF correlation functions can be written in the form
\begin{equation} \label{SF_CF}
 \langle {\cal F} \rangle  = \left\{
 {\cal Z}^{-1} \int \rmD[U]\rmD[\psi] \rmD[\psibar]\,\,
 {\cal F} \,\, \rme^{-S[U,\psibar,\psi]} 
 \right\}_{\rhoprime = \rhobarprime =\rho = \rhobar =0; C' = C =0} ,
\end{equation}
where 
${\cal F}$ stands for any product of fields localized both in the interior of
the SF box and on its time-boundaries. For instance, quark and antiquark fields at 
$x_0=0$ are given by
\begin{eqnarray}
\zeta({\bf x}) = P_- \zeta({\bf x}) & = & 
{\delta \over \delta \rhobar({\bf x}) } \, , \nonumber \\
\zetabar({\bf x}) = \zetabar({\bf x}) P_+ & = & 
-{\delta \over \delta \rho({\bf x}) } \, .
\end{eqnarray}  
The reader is referred to \cite{paper2,O(a)} for any undefined conventions.

\subsection{ Bare SF correlation functions}

Within this SF setup we now define a few on-shell correlation functions
that involve quark-antiquark pairs of boundary fields and the following
isovector composite fields:
\begin{eqnarray}  \label{quark_bilin}
A_\mu^a(x) &=& \psibar(x) \gamma_\mu \gamma_5 \half \tau^a \psi(x) \, ,
\nonumber \\
V_\mu^a(x) &=& \psibar(x) \gamma_\mu \half \tau^a \psi(x) \, ,
\nonumber \\
P^a(x)  &=& \psibar(x) \gamma_5 \half \tau^a \psi(x) \, ,
\nonumber \\
T_{\mu\nu}^a(x)  & = & \psibar(x) i \sigma_{\mu\nu}  \half \tau^a \psi(x) \, ,
\end{eqnarray}
where $\sigma_{\mu\nu} = (i/2) [\gamma_\mu,\gamma_\nu]$.

In addition to the $f$-correlators that were introduced in Ref.~\cite{paper2},
\begin{eqnarray} \label{f_corr}
  f_{\rm A}^{ab}(x_0)&=& -\langle A_0^a(x){\cal O}_{}^b\rangle, \nonumber \\[1ex]
  f_{\rm P}^{ab}(x_0)&=& -\langle P^a(x){\cal O}_{}^b\rangle, \nonumber \\[1ex]
  f_{\rm V}^{ab}(x_0)&=& -\langle V_0^a(x){\cal O}_{}^b\rangle \, ,
\end{eqnarray}
we also consider some further correlation functions:
\begin{eqnarray}  \label{k_corr}
k_{\rm A}^{ab}(x_0)&=& -\third \sum_{k=1}^3\langle A_k^a(x){\cal Q}_{k}^b\rangle, \nonumber \\[1ex]
k_{\rm T}^{ab}(x_0)&=& -\third \sum_{k=1}^3\langle T_{k0}^a(x){\cal Q}_{k}^b\rangle, \nonumber \\[1ex]
k_{\rm V}^{ab}(x_0)&=& -\third \sum_{k=1}^3\langle V_k^a(x){\cal Q}_{k}^b\rangle \, .
\end{eqnarray}
The isospin indices $a$ and $b$ are restricted to take values in the
set $\{ 1,2\}$ for reasons to be explained below, while the boundary fields 
${\cal O}_{}^a$ and ${\cal Q}_{k}^a$ are defined by:
\begin{eqnarray} \label{O-Q_k}
  {\cal O}^{a} & =  & a^6\sum_{\bfy,\bfz}
  \zetabar(\bfy)\gamma_5 \half \tau^a \zeta({\bfz}) \, ,
\nonumber \\[1ex]
  {\cal Q}_{k}^{a} & = & a^6\sum_{\bfy,\bfz}
  \zetabar(\bfy)\gamma_k \half \tau^a \zeta({\bfz}) \, .
\end{eqnarray}

For the purpose of boundary field renormalization, we also need to evaluate
a boundary-to-boundary correlator:
\begin{equation}
f_1^{ab} = - {1 \over L^6} \langle {\cal O'}_{}^a {\cal O}_{}^b \rangle,
\end{equation}
where ${\cal O'}_{}^a$ is defined analogously to ${\cal O}_{}^a$ but with
derivatives with respect to quark boundary fields at $x_0=T$ rather than
at $x_0=0$. 

\subsection{Flavour structure of the SF correlators}

As long as the isospin indices $a$ and $b$ take the values 1 or 2, 
one can show \cite{paper2} that the lattice symmetries of the tmQCD 
SF imply some exact properties of the bare correlation functions.
For the $f$-correlators  
these relations are summarized by eqs.~(I.3.48)--(I.3.49), while
for the correlator $f_1^{ab}$ one finds  
\begin{equation}
 f_1^{11} = f_1^{22} \, , \quad\quad\quad\quad f_1^{12} = f_1^{21}  = 0 \, .
\end{equation}
The analogous relations for the $k$-correlators, which can be easily derived
along the lines of Ref.~\cite{paper2}, read
\begin{equation}
   \kv^{12}(x_0)=\kt^{12}(x_0)=\ka^{11}(x_0)=0 \, ,
\end{equation}
and for ${\rm X}={\rm V,T,A}$
\begin{equation}
 \kx^{22}(x_0)=\kx^{11}(x_0),\qquad \kx^{21}(x_0)=-\kx^{12}(x_0) \, . 
\end{equation}
The non-vanishing correlators to be evaluated in practice 
can hence be chosen as $\fa^{11}(x_0)$, $\fp^{11}(x_0)$, $\fv^{12}(x_0)$,
$\kv^{11}(x_0)$, $\kt^{11}(x_0)$, $\ka^{12}(x_0)$ and $f_1^{11}$. 

An explicit representation of the $f$-correlators in terms of the
boundary-to-bulk quark propagators is given in eqs.~(I.3.50)--(I.3.52).
The analogous representations for the $k$-correlators and $f_1^{11}$ 
read\footnote{As in Ref.~\cite{LuWe96}, the bracket $\langle \dots \rangle_{\rm G}$
means an average over the gauge fields with the effective gauge action. In the
quenched approximation the average is performed with the pure gauge action.}:
\begin{eqnarray}
  \kv^{11}(x_0)&=& \half
                   \left\langle \third \sum_{k=1}^3
                   \tr\left\{
                     H_+(x)^\dagger \gamma_5\gamma_k H_+(x) \gamma_k\gamma_5
                     \right\}
                   \right\rangle_{\rm G}, \nonumber \\[1ex]
  \kt^{11}(x_0)&=&  \half
                   \left\langle \third \sum_{k=1}^3
                    \tr \left\{
                     H_+(x)^\dagger \gamma_5 \gamma_0\gamma_k H_+(x) \gamma_k\gamma_5
                       \right\}
                   \right\rangle_{\rm G}, \nonumber \\[1ex]
  \ka^{12}(x_0)&=& \frac{i}{2}
                   \left\langle \third \sum_{k=1}^3
                   \tr \left\{ 
                     H_+(x)^\dagger \gamma_k H_+(x) \gamma_k\gamma_5
                       \right\}
                   \right\rangle_{\rm G}  
\end{eqnarray}
and
\begin{equation}
  f_1^{11}  =  {\ctildet^2 \over 2} {a^6 \over L^6}
                  \sum_{{\bf y},{\bf z}} \left\langle
                  \tr \biggl. \left\{ P_+ U(y,0)^{-1} H_+(y) 
                              H_+(z)^\dagger U(z,0)
                      \right\} \biggr\vert_{y_0 = z_0 = T-a}
                  \right\rangle_{\rm G} , 
\end{equation}
where $H_+(x)$, eq.~(I.3.43), is the first flavour component of the
boundary-to-bulk quark propagator $H(x)$ defined in eq.~(I.3.38). We
remark that $f_1^{11} \geq 0$.

From the setup of the SF for tmQCD \cite{paper2} one can readily see that
evaluating $H_+$ amounts to solving for $0<x_0<T$ the one-flavour system
\begin{equation}
(D+\delta D+m_0+i\muq\gamma_5) \widetilde{H}_+(x)
= \ctildet a^{-1} \delta_{x_0,a} U(x-a\hat{0},0)^{-1} P_{+} ,
  \label{propagator}
\end{equation}
with the boundary conditions
\begin{equation}
P_{+} \widetilde{H}_+(x) \bigl\vert_{x_0=0} \; = \; 
P_{-} \widetilde{H}_+(x) \bigl\vert_{x_0=T} \; = \; 0 .
\end{equation}
The solution $\widetilde{H}_+(x)$ of eq.~(\ref{propagator}) satisfies
$\widetilde{H}_+(x) P_+ = \widetilde{H}_+(x)$ and is trivially related
to the boundary-to-bulk quark propagator $H_+(x)$:
\begin{equation}
\widetilde{H}_+(x) = H_+(x) - \delta_{x_0,0} P_+  \, .
\end{equation}

Provided the triplet isospin indices are restricted to the values 1 and 2,
we are able to express 
the correlation functions of Subsection 2.1 in terms of $H_+(x)$ alone,   
which saves about a factor of two in CPU-time. Since
the full physical isospin symmetry is expected to be restored in the 
continuum limit of lattice tmQCD \cite{paper1}, the above restriction
implies no loss of physical information.

\section{Renormalization scheme and scaling observables}
The SF for lattice tmQCD is expected to be ultraviolet finite after
renormalization of the bare parameters in the action, $g_0^2$, $m_0$
and $\muq$, and the boundary quark fields \cite{paper2}. As the latter renormalize
multiplicatively and are set to zero, here we do not have to worry 
about their renormalization. In the following we specify our renormalization
scheme for the parameters in the action and the correlation functions. 
Provided that all the relevant improvement coefficients are given their
proper values, the mutual relations among renormalized parameters and 
observables are free from O($a$) cutoff effects.

\subsection{Renormalized parameters}
Since we work in the quenched approximation to QCD, it is convenient 
to renormalize the gauge coupling by eliminating $g_0^2$ in favour of the
hadronic length scale $r_0 \simeq 0.5 \, {\rm fm}$ \cite{r0intro} and then
express all the physical quantities in units of $r_0$.  
In large physical volume the 
relation between $\beta=6/g_0^2$ and $a/r_0$ has been evaluated
\cite{r0scale} with a relative accuracy of about $0.5\%$. 
Since for the present
scaling study we are working in an intermediate volume, where finite-size
effects are non-negligible, we keep constant the ratio $L/r_0$,
\begin{equation} \label{L_over_r0-beta}
{ L \over r_0} = \left[{ a \over r_0}\right](\beta) \, { L \over a }  =  1.49 \, ,
\end{equation}
while approaching the continuum limit.
We choose $T/a =2L/a$ and values of $L/a$ such that the values of $\beta$
lie in the range $6 \leq \beta \leq 6.5$, which is the one of interest 
for quenched lattice QCD with the Wilson plaquette action.

As for the renormalization of the quark mass parameters, we require
\begin{eqnarray} \label{mass_rencond}
L \mr & = & 0.020 \, , \nonumber  \\[1ex]
L \mur & = & 0.153 \, ,
\end{eqnarray} 
where $\mr$ and $\mur$ are the renormalized quark mass parameters
introduced in eqs.~(I.2.8)--(I.2.9). Following Ref. \cite{paper2},
the renormalized twisted mass parameter $\mur$ is related to the bare
masses $\muq$ and $\mq = m_0 - \mc$ via
\begin{equation} \label{mu_R_def}
\mur = Z_\mu (1+\bmu a\mq) \muq \, , 
\end{equation} 
where $\bmu$ is an improvement coefficient introduced in \cite{paper2}.
The exact lattice PCVC relation (I.2.14) implies that we can set:
\begin{equation} \label{Z_mu_def}
Z_\mu = Z_{\rm P}^{-1} \, .
\end{equation}
The renormalization constant of the isotriplet pseudoscalar density $Z_{\rm P}$
is evaluated in the SF scheme at the momentum scale $q=(1.436 r_0)^{-1}$, using
the results of Ref.~\cite{ZP}.  
The value of $\mr$ in the same scheme and at the same scale is computed
from the renormalized PCAC relation as discussed later on.

We recall from Ref.~\cite{paper1}
that in renormalized tmQCD the "polar" quark mass 
\begin{equation}
  \Mr \equiv \sqrt{\mr^2 + \mur^2} 
\end{equation}
plays the role of the renormalized quark mass.
The angle $\alpha$, defined by
\begin{equation}
\tan \alpha \equiv {\mur \over \mr} \, ,
\end{equation}
can be chosen arbitrarily and just determines the physical interpretation
of the tmQCD correlation functions. The numerical values on the r.h.s.
of eq.~(\ref{mass_rencond}) yield $L \Mr \simeq 0.154$,
which is in the range of the scaling study \cite{Jo}, and a value of $\alpha$   
that is far from zero, namely $\pi/2-\alpha  \simeq  0.130$.

\subsection{Renormalized and O($a$) improved correlators}

The definition of renormalized and O($a$) improved SF correlators
is guided by the form of the renormalized and O($a$) improved bulk
fields, eqs.~(I.2.10)--(I.2.12) and
\begin{equation} \label{biquark_tens_RI}
(T_{\rmR})^a_{\mu\nu} = Z_{\rm{T}} (1+b_{\rm{T}} a \mq)
[ T_{\mu\nu}^a+c_{\rm{T}}a
(\tilde{\partial}_{\mu}V_{\nu}^a-\tilde{\partial}_{\nu}V_{\mu}^a) ] \, ,
\end{equation}
where $\tilde{\partial}_{\mu}$ denotes the symmetric lattice derivative in
the direction of the unit vector $\hat{\mu}$. We hence define:
\begin{eqnarray} \label{ren_SFcorr} \nonumber
[\fa^{11}(x_0)]_{\hbox{\sixrm R}} &\!\! = \!\!& 
                [Z_\zeta (1+ b_\zeta a\mq)]^2 Z_{\rm{A}} (1+b_{\rm{A}} a \mq)
                \left[ \fa^{11} + c_{\rm A} a\tilde{\partial}_0 \fp^{11}
                - a\muq\,\tba\, \fv^{12}
                \right](x_0) \, , \\[1ex] \nonumber
[\fv^{12}(x_0)]_{\hbox{\sixrm R}} &\!\! = \!\!& 
                [Z_\zeta (1+ b_\zeta a\mq)]^2 Z_{\rm{V}} (1+b_{\rm{V}} a \mq)
                \left[ \fv^{12}
                + a\muq\,\tbv\, \fa^{11}
                \right](x_0) \, , \\[1ex] \nonumber
[\fp^{11}(x_0)]_{\hbox{\sixrm R}} &\!\! = \!\!& 
                [Z_\zeta (1+ b_\zeta a\mq)]^2 Z_{\rm{P}} (1+b_{\rm{P}} a \mq)
                \, \fp^{11} (x_0) \, , \\[1ex] \nonumber
[\kv^{11}(x_0)]_{\hbox{\sixrm R}} &\!\! = \!\!& 
                [Z_\zeta (1+ b_\zeta a\mq)]^2 Z_{\rm{V}} (1+b_{\rm{V}} a \mq)
                \left[ \kv^{11} + c_{\rm V} a\tilde{\partial}_0 \kt^{11}
                - a\muq\,\tbv\, \ka^{12}
                \right](x_0) \, , \\[1ex] \nonumber
[\kt^{11}(x_0)]_{\hbox{\sixrm R}} &\!\! = \!\!& 
                [Z_\zeta (1+ b_\zeta a\mq)]^2 Z_{\rm{T}} (1+b_{\rm{T}} a \mq)
                \left[ \kt^{11} - c_{\rm T} a\tilde{\partial}_0 \kv^{11}
                \right](x_0) \, , \\[1ex] 
[\ka^{12}(x_0)]_{\hbox{\sixrm R}} &\!\! = \!\!& 
                [Z_\zeta (1+ b_\zeta a\mq)]^2 Z_{\rm{A}} (1+b_{\rm{A}} a \mq)
                \left[ \ka^{12}
                + a\muq\,\tba\, \kv^{11}
                \right](x_0)   
\end{eqnarray}
and 
\begin{equation}
[f_1^{11}]_{\hbox{\sixrm R}} \; = \; [Z_\zeta (1+ b_\zeta a\mq)]^4 f_1^{11} \, .
\end{equation}
The improvement coefficients $\tba$ and $\tbv$ have been introduced in \cite{paper2},
whereas all the remaining improvement coefficients are the same as
in lattice QCD with standard quark mass
parameterization. We remark that the expressions for
$[\fv^{12}]_{\hbox{\sixrm R}}$ and $[\ka^{12}]_{\hbox{\sixrm R}}$ 
are independent of $c_{\rm V}$ and $c_{\rm A}$, respectively, because of the
translational invariance of the theory in the spatial directions.

\subsection{Pion and $\rho$-meson channel correlators}

The definition of the observables for this scaling test is inspired by the criterion
of considering observables that in the limit of large $T$ and large $L$ turn into 
the pion and $\rho$-meson mass and decay constant, except for the normalization
of the $\rho$-meson decay constant which is not the physical one. The same criterion
was followed in the scaling study of Ref.~\cite{Jo}. 

The first step in the construction of the meson observables is to build linear
combinations of the correlators in eq.~(\ref{ren_SFcorr}) so to yield at time
$x_0$ insertions of operators with the appropriate quantum numbers 
to create/annihilate a pion or a $\rho$-meson (or higher states in the
same channels). According to the relation between renormalized correlation
functions of QCD and tmQCD in infinite volume \cite{paper1}, such operators
can be written as follows:
\begin{eqnarray} \label{oper_tmbasis}
(A'_{\rmR})^a_0(x) & = & \cos \alpha (A_{\rmR})^a_0(x) + 
       \varepsilon^{3ac} \sin \alpha (V_{\rmR})^c_0(x)  \, ,
\nonumber \\[1ex] 
(P'_{\rmR})^a(x) & = & (P_{\rmR})^a(x) \, ,
\nonumber \\[1ex] 
(V'_{\rmR})^a_k(x) & = & \cos \alpha (V_{\rmR})^a_k(x) +
       \varepsilon^{3ac} \sin \alpha (A_{\rmR})^c_k(x) \, , 
\nonumber \\[1ex] 
(T'_{\rmR})^a_{k0}(x) & = & (T_{\rmR})^a_{k0}(x) \, .
\end{eqnarray}
We remark that the expression of local operators with given physical quantum numbers
in the tmQCD quark basis does not depend on the choice of boundary conditions,
and the results of Ref.~\cite{paper1} are hence valid in the present context. The
situation is different for the correlation function themselves, so that the SF
correlators defined below cannot be directly compared --at least for finite extent $T$
of the SF-- with those computed at the same value of $\Mr$ and $\alpha = 0$. 
A detailed discussion of this point is deferred to a forthcoming publication \cite{paper4}.

The correlators containing the operator insertions in eq.~(\ref{oper_tmbasis}) with
isospin index $a=1$ are correspondingly given by 
\begin{eqnarray} \label{pion_rho_corr}
[\fap^{11}(x_0)]_{\hbox{\sixrm R}} & = & 
\cos \alpha [\fa^{11}(x_0)]_{\hbox{\sixrm R}} 
-  \sin \alpha [\fv^{12}(x_0)]_{\hbox{\sixrm R}} \, ,
 \nonumber \\[1ex] 
[\fpp^{11}(x_0)]_{\hbox{\sixrm R}} & = & [\fp^{11}(x_0)]_{\hbox{\sixrm R}} \, ,  
 \nonumber \\[1ex]
[\kvp^{11}(x_0)]_{\hbox{\sixrm R}} & = & 
\cos \alpha [\kv^{11}(x_0)]_{\hbox{\sixrm R}} 
- \sin \alpha [\ka^{12}(x_0)]_{\hbox{\sixrm R}} \, ,
\nonumber \\[1ex] 
[\ktp^{11}(x_0)]_{\hbox{\sixrm R}} & = & [\kt^{11}(x_0)]_{\hbox{\sixrm R}} \, .
\end{eqnarray}
With the SF-boundary fields ${\cal O}^1$ or ${\cal Q}_k^1$ introduced in Subsection~2.1,
one expects that in the limit of large $x_0$ and 
$T-x_0$ the correlators in eq.~(\ref{pion_rho_corr})
are dominated by the "$\!$pion" and "$\!\rho$-meson" states in spatial volume $L^3$,
respectively.

\subsection{The observables of this scaling test}

We are now ready to define the scaling observables which we will focus on in the
remaining part of this paper. 

\begin{itemize}
\item {\bf Meson observables} \\
In terms of the above correlators, eq.~(\ref{pion_rho_corr}), the estimators of
the finite volume pion (PS) and $\rho$-meson (V) masses read: 
\begin{eqnarray} \label{masses_PS_V}
m_{\rm{PS}}= - 
\frac{\tilde{\partial}_0 [\fpp^{11}]_{\hbox{\sixrm R}}}{[\fpp^{11}]_{\hbox{\sixrm R}}}
\Biggr|_{x_0=T/2} \, ,
& \quad\quad &
\widetilde{m}_{\rm {PS}}= - 
\frac{\tilde{\partial}_0 [\fap^{11}]_{\hbox{\sixrm R}}}{[\fap^{11}]_{\hbox{\sixrm R}}}
\Biggr|_{x_0=T/2} \, ,
\nonumber  \\
m_{\rm{V}}= - 
\frac{\tilde{\partial}_0 [\kvp^{11}]_{\hbox{\sixrm R}}}{[\kvp^{11}]_{\hbox{\sixrm R}}}
\Biggr|_{x_0=T/2} \, ,
& \quad\quad &
\widetilde{m}_{\rm{V}}= - 
\frac{\tilde{\partial}_0 [\ktp^{11}]_{\hbox{\sixrm R}}}{[\ktp^{11}]_{\hbox{\sixrm R}}}
\Biggr|_{x_0=T/2} \, .
\end{eqnarray}
It should be noted that at finite $T$ the quantities $m_{\rm{PS}}$ and
$m_{\rm{V}}$ need not coincide with $\widetilde{m}_{\rm {PS}}$ and
$\widetilde{m}_{\rm{V}}$, respectively, because they may receive contributions
from states heavier than the finite volume pion or $\rho$-meson.

The estimators of the finite volume pion and $\rho$-meson decay constants read
\begin{eqnarray} \label{etas_PS_V}
\eta_{\rm{PS}} & =  &  [f_1^{11}]_{\hbox{\sixrm R}}^{-1/2} \;
C_{\rm{PS}} \; [ \fap^{11}(x_0) ]_{\hbox{\sixrm R}}\Bigr|_{x_0=T/2} \,  , 
\nonumber \\
\widetilde{\eta}_{\rm{PS}} & =  & [f_1^{11}]_{\hbox{\sixrm R}}^{-1/2} \;
\widetilde{C}_{\rm{PS}} \; [ \fap^{11}(x_0) ]_{\hbox{\sixrm R}}\Bigr|_{x_0=T/2} \, ,
\nonumber \\
\eta_{\rm{V}} & =  & [f_1^{11}]_{\hbox{\sixrm R}}^{-1/2} \;
C_{\rm{V}} \; [ \kvp^{11}(x_0) ]_{\hbox{\sixrm R}}\Bigr|_{x_0=T/2} \, ,
\nonumber \\
\widetilde{\eta}_{\rm{V}} & =  & [f_1^{11}]_{\hbox{\sixrm R}}^{-1/2} \;
\widetilde{C}_{\rm{V}} \; [ \kvp^{11}(x_0) ]_{\hbox{\sixrm R}}\Bigr|_{x_0=T/2} \, ,
\end{eqnarray}
where the normalization constants $C_{\rm{PS}}$ and $C_{\rm{V}}$ are given by
\begin{equation}
C_{\rm{PS}} = \frac{2}{\sqrt{ L^3 m_{\rm{PS}}^{\phantom{3}} }} \, , \quad\quad\quad
C_{\rm{V}} = \frac{2}{\sqrt{ L^3 m^3_{\rm{V}} }} \, .
\end{equation}
The constants $\widetilde{C}_{\rm{PS}}$ and $\widetilde{C}_{\rm{V}}$ are defined
analogously in terms of $\widetilde{m}_{\rm{PS}}$ and $\widetilde{m}_{\rm{V}}$.
The normalization constant $C_{\rm{PS}}$ is chosen such that $\eta_{\rm{PS}}\to F_\pi$ 
as $T=2L\to\infty$. In the same limit the quantity $\eta_{\rm{V}}$, due to its 
unphysical normalization, does not approach the (inverse) decay constant of the $\rho$-meson,
but one may expect the cutoff effects to be similar to those of the properly normalized
estimator of $1/F_\rho$.
Analogous remarks hold for the quantities $\widetilde{\eta}_{\rm{PS}}$
and $\widetilde{\eta}_{\rm{V}}$, which differ from $\eta_{\rm{PS}}$
and $\eta_{\rm{V}}$ only by their normalization. 

\item {\bf PCVC and PCAC quark masses} \\
The PCVC and PCAC operator relations of renormalized tmQCD, which follow from
the flavour chiral Ward identities \cite{paper1}, imply corresponding relations 
among the renormalized SF correlators introduced above:
\begin{equation} \label{PCVC}
\tilde{\partial}_0 [\fv^{12}(x_0)]_{\hbox{\sixrm R}}  =  
-2 \mur [\fp^{11}(x_0)]_{\hbox{\sixrm R}}
\end{equation}
and
\begin{equation} \label{PCAC}
\tilde{\partial}_0 [\fa^{11}(x_0)]_{\hbox{\sixrm R}}  =
 2 \mr [\fp^{11}(x_0)]_{\hbox{\sixrm R}}  \, .
\end{equation}
As a consequence of the improvement of the bulk action and relevant operators,
at finite lattice spacing $a$ the cutoff effects on these relations are O($a^2$), 
even without improving the SF-boundary action and fields.
  
A way of checking the size of the residual cutoff effects in the PCVC relation,
eq.~(\ref{PCVC}), is to consider the quantity
\begin{equation} \label{r_PCVC_def}
r_{\rm PCVC} = { \muq \over \bar{\mu} } \, ,
\end{equation}
where $\bar{\mu}$ is the estimate of the bare current twisted mass obtained
from the SF correlators: 
\begin{equation} \label{mubar_def}
\bar{\mu} = - Z_{\rm V} (1+b_{\rm V}a\mq)
            \left. {\tilde{\partial}_0 \fv^{12}(x_0) \over 2 \fp^{11}(x_0) }
            \right|_{x_0 = T/2} \, .
\end{equation}
In the definition of $\bar{\mu}$ we have left out the improvement
coefficients that are only perturbatively known\footnote{However,
a non-perturbative estimate of $b_{\rm P}$ at $\beta=6$ and $\beta=6.2$ 
has been given in Ref.~\cite{gupta}.}.
Close to the continuum limit we expect:
\begin{equation} \label{r_PCVC_a_eff}
r_{\rm PCVC} = 1 - a \mr Z_{\rm m}^{-1} [ b_{\rm P} +
               b_\mu + Z Z_{\rm V} \tbv] + {\rm O}(a^2) \, ,
\end{equation}
where $Z = Z_{\rm m} Z_{\rm P}/Z_{\rm A}$ (see e.g. Ref.~\cite{bambp} for a
recent non-perturbative estimate of $Z$ as a function of $g_0^2$).
In view of the very small values of $a \mr$ that correspond to $L/a \geq 8$ and 
$L\mr = 0.020$,
the cutoff effects on $r_{\rm PCVC}$ should be dominated by the terms O($a^2$),
and the sensitivity to the combination $[ b_{\rm P} + b_\mu + Z Z_{\rm V} \tbv]
\simeq {\rm O}(1)$ should hence be very small. Our data (see Section 4) confirm this
expectation.

The PCAC relation, eq.~(\ref{PCAC}), is instead exploited to evaluate the renormalized
standard mass, which is defined by 
\begin{equation} \label{mr_def}
\mr \equiv
{ \tilde{\partial}_0 [\fa^{11}(x_0)]_{\hbox{\sixrm R}} \over
2 [\fp^{11}(x_0)]_{\hbox{\sixrm R}} }\Biggr|_{x_0 = T/2} \, .
\end{equation}
\end{itemize}
One could of course evaluate $\mr$ by eq.~(I.2.5) and eq.~(I.2.8), but then
the absolute accuracy on $L\mr$ is essentially limited by the accuracy on 
$\mc = \mc(g_0^2)$. Turning the argument around, one can exploit the determination
of $\mr$ obtained from eq.~(\ref{mr_def}) to estimate the critical quark mass $\mc$,
which is independent of $\alpha$ up to cutoff effects \cite{paper1}. Within the tmQCD
regularization, the estimate of $\mc$ can be performed for any $\beta$ with no
singularities in the computation of quark propagators by working at very small
values of $\mr$ and reasonable finite values of $\mur$. Moreover from Ref.~\cite{paper2}
it can be argued that an O($a$) 
improved evaluation of $\mc$ in lattice tmQCD requires the non-perturbative knowledge
of a certain combination of the improvement coefficients $ \tba$ and $ \tbm$: 
$$\tbm - (Z Z_{\rm V})^{-1} \tba \, .$$
Work in this direction is currently in progress by the Tor Vergata APE 
group \cite{tov}.

\section{Numerical details and results}

The basic idea of any scaling test is to approach the continuum limit along a
line in bare parameter space where all renormalized parameters are kept constant,
which is achieved here by the renormalization conditions specified in Subsection 3.1.
Under these conditions the renormalized and (almost) O($a$) improved observables 
that we introduced in Subsection 3.4 are expected to depend on $a/L$ only and 
converge to a well-defined continuum limit as $a/L \to 0$ with (almost) no scaling
violations linear in $a/L$.

\subsection{Renormalization constants and improvement coefficients}
The renormalization conditions for the gauge coupling, 
eq.~(\ref{L_over_r0-beta}), and for the two quark mass parameters,
eq.~(\ref{mass_rencond}), where the scheme dependence arises only
from $Z_{\rm P}$, are sufficient to render ultraviolet
finite the observables introduced in Subsection 3.4. Indeed, no
observables depend on the boundary field renormalization factor 
$Z_\zeta (1+ b_\zeta a\mq)$, as well as on the product
$Z_{\rm T} (1+ b_{\rm T} a\mq)$. Concerning $\za$ and $\zv$,
which are needed to remove lattice artifacts that vanish more
slowly than $a$ as $a \to 0$, we employ the available non-perturbative
estimates from Ref.~\cite{O(a)2}. 
Since all these renormalization constants are defined either
at the chiral point of quenched QCD or in the Yang-Mills
SU(3) theory, we have actually set up a non-perturbative 
quark mass independent scheme.

In order to further reduce the scaling violations of our observables, we have
to give proper values to all the relevant improvement coefficients.
In the limit of large time extent $T$ of the SF system, the O($a$)
improvement of the bulk action and operators is sufficient to improve
the scaling observables defined in Subsection 3.4. As we actually
work at $T \simeq 1.5$ fm, this statement remains valid only for the
quantity $r_{\rm PCVC}$, eq.~(\ref{r_PCVC_def}). The remaining scaling
observables, which are defined at $x_0 = T/2$, are indeed not completely
dominated by the pion or the $\rho$-meson state and hence do still 
depend on the details of the SF-boundary fields. As a consequence, the
O($a$) improvement of the SF boundary action and fields can not be neglected.

Let us start with the coefficients that are relevant for the improvement
of the massless theory. We employ the non-perturbative estimates
of $\csw$, $\ca$ and $\cv$ computed in Refs.~\cite{NP_O(a),cV_proc},
while setting $c_{\rm T}$ to its one-loop value\footnote{We recall that 
$c_{\rm T}$ is only relevant for the improvement of $\widetilde{m}_{\rm{V}}$ 
and $\widetilde{\eta}_{\rm{V}}$.}~\cite{O(a)pert}. As for the improvement 
of the SF-boundaries, the only coefficients relevant for our study are
$\ct$ and $\ctildet$, which are both set to their one-loop values
\cite{SFYM,LuWe96}. We discuss below the impact of reasonable changes
of these values on our scaling observables. 

In view of the very small values of $a\mr$ corresponding to $L \mr = 0.020$,
the improvement coefficients multiplying counterterms that are linear in
$a\mr$ need not be very precisely tuned. Nevertheless, we adopted non-perturbative
estimates of $b_{\rm A}-b_{\rm P}$ \cite{bambp} and $b_{\rm V}$ \cite{O(a)pert},
and one-loop estimates for $b_{\rm A}$ and $b_{\rm P}$ \cite{O(a)pert}, whereas 
$b_{\rm T}$ is not necessary at all for the improvement of our observables.
It should be noted that, due to the quark mass renormalization 
conditions~(\ref{mass_rencond}), which entail $\alpha \simeq \pi/2$,
the correlator $[\fap^{11}(x_0)]_{\hbox{\sixrm R}}$ depends on $b_{\rm A}$ only
via a term proportional to $a\mr \cos \alpha \ll 1$. For the same reason,
the ratio 
$\tilde{\partial}_0 [\kvp^{11}]_{\hbox{\sixrm R}} / [\kvp^{11}]_{\hbox{\sixrm R}}$
and thus $m_{\rm{V}}$ are almost independent of $b_{\rm A}$.  

Among the improvement coefficients that multiply counterterms of order
$a\mur$ \cite{paper2}, in this scaling test we only need to properly tune
$\tba$, eq.~(I.2.10), $\tbv$, eq.~(I.2.11), and $\tilde{b}_1$, eq.~(I.3.6).
Moreover, since the set of improvement coefficients $\{\,\tilde{b}_1,
\tilde{b}_{\rm{m}},b_{\mu},\tilde{b}_{\rm{A}},\tilde{b}_{\rm{V}}\}$
is actually redundant, as explained in Ref.~\cite{paper2}, one of them
can be arbitrarily prescribed. For practical reasons that are specific to
this scaling test, we find it convenient to set $\tilde{b}_1(g_0^2) \equiv 1$
exactly. We can then completely
forget about the corresponding O($a$) boundary counterterms. Concerning
$\tba$, $\tbv$ and $\bmu$, they are set to the one-loop values \cite{paper2} that
follow from the above prescription of $\tilde{b}_1$, i.e.  
\begin{equation} \label{1l_tildeb}
\tba = 0.0213 g_0^2 , \quad\quad\quad \tbv = 0.0053 g_0^2 ,
\quad\quad\quad \bmu = -0.0440 g_0^2 \, . 
\end{equation}
We remark that the counterterm proportional to $\bmu$ is of order $a\mr$, while 
$\tbm$ is not needed at all for this study. 
It remains to be checked a posteriori whether the residual O($a\mur$) effects
are significant in comparison to the higher order scaling
violations and the statistical uncertainties of our observables.





\subsection{Simulation parameters and analysis of the raw data}
In order to check the size of the scaling violations in our observables and the 
rate of the approach to the continuum limit, we perform simulations with
four different lattice resolutions in the range $2 {\rm GeV} \leq 1/a \leq
4 {\rm GeV}$, while enforcing the renormalization conditions detailed above.
Throughout the whole analysis we adopt the definitions of the renormalization
constants and the improvement coefficients specified in Subsection 4.1 and
evaluate them at the values of $\beta=6/g_0^2$ chosen for our simulations.

An overview of the bare parameters, the corresponding renormalized parameters and
the accumulated statistics is given in Table~\ref{tab:sim1}. The values of $L/r_0$ 
and $L\mur$ follow from our choice of bare parameters, and the quoted 
uncertainties stem from the statistical errors on $a/r_0$ \cite{r0scale}
and on $\zp=\zp(q)$ \cite{ZP}, respectively.
The uncertainty on $L\mr$ reflects instead the statistical error on the
SF correlators in eq.~(\ref{mr_def}), which in turn is a combination of 
the statistical errors on the bare SF correlators, as evaluated via simulations
of lattice tmQCD, and the known ratio $\za/\zp$.
\begin{table}[htb]
\vspace{0.25cm}
\begin{center}
\begin{tabular}{ccccccccc}
\hline
  set & $L/a$ & $\beta$ & $a\muq$ & $\kappa$ & $L/r_0$
& $L\mur$ & $L\mr$ & $N_{\rm meas}$ \\
\hline
  A &  8 & 6.0  & 0.01    & 0.134952 & 1.490(6) & 0.1529(8)
& 0.0228(23) & 7680 \\
  B & 10 & 6.14 & 0.00794 & 0.135614 & 1.486(7) & 0.1530(8)
& 0.0203(30) & 2880 \\
  C & 12 & 6.26 & 0.00659 & 0.135742 & 1.495(7) & 0.1530(8)
& 0.0201(23) & 3072 \\
  D & 16 & 6.47 & 0.00493 & 0.135611 & 1.488(7) & 0.1529(8)
& 0.0180(24) & 1680 \\
\hline
\end{tabular}
\caption{{\sl
The bare and renormalized parameters for our data sets
and the number ($N_{\rm meas}$) of computed SF quark propagators on decorrelated
gauge backgrounds.
}}\label{tab:sim1}
\end{center}
\end{table}
Our Monte Carlo simulations of quenched lattice tmQCD are performed on the APE100
parallel computers with 32--256 nodes at INFN Milan and DESY Zeuthen.
The parallelization of the program and the machine topology allow us to
simulate several independent replica of the smaller systems (i.e.~A,B and
C in Table~\ref{tab:sim1}) at the same time. The computational effort needed
for the present scaling test amounts to about 75 Gflops $\times$ days.

The gauge configurations are generated using a standard hybrid overrelaxation
algorithm. The single iteration is defined by one heatbath step followed by
$N_{\rm OR}=L/2a +1$ microcanonical reflection steps.
The correlation functions are evaluated by averaging over sequential gauge
field configurations separated by 50 iterations.
For the computation of the quark propagators $H_+(x)$ entering our observables, we
use the BiCGStab inversion algorithm with SSOR preconditioning \cite{SSOR}.
As a stopping criterion for the inversion algorithm we require the square
norm of the dynamical residue, as defined in Ref.~\cite{SSOR}, to be
$10^{13}$ times smaller than the square norm of the solution. For the finest
lattice spacing considered, convergence is always reached within a number of
BiCGStab iterations between 80 and 120.

A binning analysis of our data shows that for all simulation points
consecutive measurements of the correlation functions (in the above specified sense)
can be effectively taken as statistically independent.
As our observables are non-linear combination of the basic correlation
functions, we adopt a single-elimination jackknife procedure for the
evaluation of their statistical errors.

\subsection{Continuum limit extrapolations}

\begin{figure}[htb]
\begin{center}
\epsfig{file=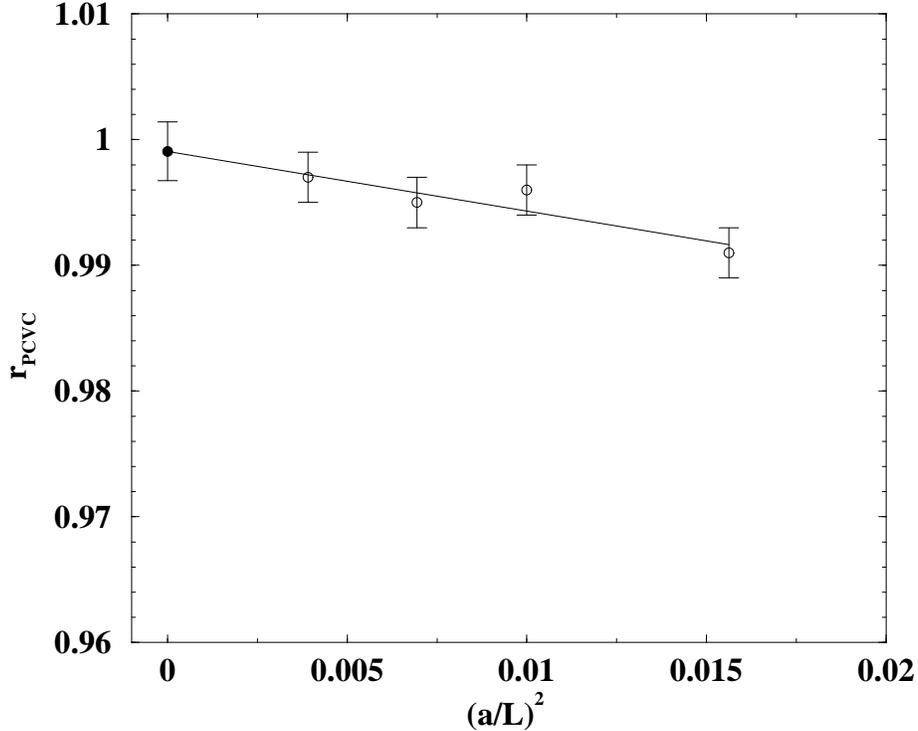,width=10.0cm,angle=-90}
\caption{{\sl
Our results for $r_{\rm PCVC}$ and their continuum limit, obtained
by fitting the four data points to a first order polynomial in $(a/L)^2$.
}}\label{fig:r_PCVC}
\end{center}
\end{figure}

By the above analysis procedure we obtain the results for $r_{\rm PCVC}$ shown 
in Figure~\ref{fig:r_PCVC} and the values of the meson observables quoted 
in Table~\ref{tab:jo2}. 

The errors on our results for $r_{\rm PCVC}$ already take into
account the small statistical uncertainty\footnote{We neglect here
the tiny systematic uncertainties associated to the determination of
$\zv$: see Ref.~\cite{O(a)2} for details.} on $\zv$, whereas the product
$\bv a \mq \ll 1$ is taken with no error. Inspection of Figure~\ref{fig:r_PCVC}
immediately reveals that the residual cutoff effects of order $a\mr$
--see eq.~(\ref{r_PCVC_a_eff})-- are completely negligible with respect to 
the very small statistical errors and higher order scaling violations.
The very tiny mismatch between the values of $L\mr$ and the nominal
value $0.020$ is hence completely negligible in this case. Moreover, 
due to its very definition, $r_{\rm PCVC}$ is independent of $\zp$ 
and $\cv$.
We hence conclude that for $\alpha \simeq \pi/2$ and the values of $a\muq$
that are relevant for (quenched) QCD in the chiral regime, the PCVC relation 
shows surprisingly small cutoff effects and is effectively O($a$)
improved once the proper value of $\csw$ is employed.  
\begin{table}[htb]
\vspace{0.25cm}
\begin{center}
\begin{tabular}{ccccc}
\hline
set &  $m_{\rm PS}L$ & $m_{\rm V}L$ & $\eta_{\rm PS}L$ & $\eta_{\rm V}$ \\
\hline
 A   & 1.866(14)      & 2.693(18)      & 0.5419(34)          & 0.1528(21)
\\
 B   & 1.805(21)      & 2.652(27)      & 0.5570(56)          & 0.1565(34)
\\
 C   & 1.831(21)      & 2.646(28)      & 0.5514(53)          & 0.1615(36)
\\
 D   & 1.825(27)      & 2.648(35)      & 0.5510(70)          & 0.1601(46)
\\
\hline
set  &  $\widetilde{m}_{\rm PS}L$ & $\widetilde{m}_{\rm V}L$
& $\widetilde{\eta}_{\rm PS}L$ & $\widetilde{\eta}_{\rm V}$        \\
\hline
 A   & 1.713(8)       & 2.398(12)      & 0.5654(25)          & 0.1817(21)
\\
 B   & 1.667(11)      & 2.337(19)      & 0.5793(38)          & 0.1889(34)
\\
 C   & 1.680(10)      & 2.345(19)      & 0.5751(36)          & 0.1933(34)
\\
 D   & 1.659(13)      & 2.323(25)      & 0.5779(49)          & 0.1949(47)
\\
\hline
\end{tabular}
\caption{{\sl "Raw" results for our meson observables: the quoted errors arise
just from statistical fluctuations over our samples of gauge configurations.
}}
\label{tab:jo2}
\end{center}
\end{table}

In the case of our meson observables, which may be quite sensitive
to mismatches and uncertainties in the renormalization conditions as well
as to uncancelled O($a$) cutoff effects stemming from the SF-boundaries,
we have performed a slightly more refined analysis before producing scaling
plots and attempting continuum extrapolations. The starting point of this
further analysis is represented by Table~\ref{tab:jo2}, which is directly
obtained from the simulation data by considering all the renormalization
constants and improvement coefficients with no error. The quoted errors arise
just from the statistical fluctuations of the observables over the samples
of gauge configurations produced at the bare parameters of Table~\ref{tab:sim1}.
The remaining uncertainties are taken into account as follows.

\begin{itemize}

\item {\bf Uncertainties on the renormalization and improvement coefficients} \\
The observables $\widetilde{m}_{\rm {PS}}$ and $m_{\rm V}$ have a very tiny
dependence on $\za/\zv$ which disappears in the limit $\alpha \to \pi/2$.
Since $\alpha \simeq \pi/2$, the would-be decay constants of the pion
and the $\rho$-meson are almost proportional to $\zv$ and $\za$, respectively.
The statistical uncertainties on $\zv$ and $\za$ \cite{O(a)2}, which are of about  
0.01\% and 1\% of the mean values, respectively, are 
added quadratically to the errors in Table~\ref{tab:jo2}.
The systematic uncertainties on $\zv$ and $\za$ \cite{O(a)2} are shown separately as 
errors on the continuum limit extrapolations, see Table~\ref{tab:dev}.

The uncertainties on the improvement coefficients that are needed to
subtract effects of order $a\muq$ or $a\mq$ can safely be neglected,
as we work with $a\muq \leq 0.01$ and $a\mq$ one order of magnitude 
smaller than $a\muq$. Concerning the uncertainties on $\ca$, $\cv$     
and $\cT$, their statistical errors represent very tiny effects,
whereas the intrinsic O($a$) ambiguity of these coefficients by definition 
affects any scaling observables only at O($a^2$). Nevertheless, in the case 
of $\cv$, for which the available non-perturbative estimates at low values
of $\beta$ are non-small (of order 0.1) and significantly different from 
each other, one might want to check\footnote{We thank the referee  
for suggesting this check.} the effect of
employing for our observables the one-loop value \cite{O(a)pert}
or the non-perturbative estimate by Ref.~\cite{gupta}
rather than the value determined in Ref.~\cite{cV_proc}.

By definition, among our observables, only 
$m_{\rm V}L$, $\eta_{\rm V}$ and $\wt{\eta}_{\rm V}$
depend on $\cv$. The dependence on $\cv$ is due to the
contribution from $[\kv^{11}(x_0)]_{\hbox{\sixrm R}}$
to the correlator $[\kvp^{11}(x_0)]_{\hbox{\sixrm R}}$. Since this term
comes with a factor of $\cos \alpha$, see eq.~(\ref{pion_rho_corr}), one can
expect that for our choice of renormalized quark mass parameters, 
eq.~(\ref{mass_rencond}), the $\cv$-dependent contribution to 
$[\kvp^{11}(x_0)]_{\hbox{\sixrm R}}$ is quite small. 
Indeed, at $\beta=6$ we find   
\be
m_{\rm V}L = 2.697(18) \; , \quad\quad\quad
\eta_{\rm V} = 0.1547(21) \; , \quad\quad\quad
\wt{\eta}_{\rm V} = 0.1845(21)
\ee
with the one-loop value of $\cv$ and
\be
m_{\rm V}L = 2.696(18) \; , \quad\quad\quad
\eta_{\rm V} = 0.1541(21) \; , \quad\quad\quad
\wt{\eta}_{\rm V} = 0.1835(21)
\ee
with the non-perturbative value of $\cv$ given in Ref.~\cite{gupta}.
Comparing with the values quoted in Table~\ref{tab:jo2} (set A), we see that
on $m_{\rm V}L$ the effect of using the value of $\cv$ by Ref.~\cite{gupta}
is negligibly small, while the analogous effect on $\eta_{\rm V}$ 
and $\wt{\eta}_{\rm V}$ is less than one standard deviation. Moreover,
employing the one-loop value of $\cv$ at $\beta=6.26$ we observe deviations
from the results of Table~\ref{tab:jo2} that are smaller by about a factor 
of two than the corresponding deviations found at $\beta=6$.
We hence conclude that the choice of the non-perturbative definition of $\cv$ 
is not important for the scaling behaviour of our observables. 

The available non-perturbative estimates
of $\ca$ are smaller than the corresponding ones of $\cv$ by at least a factor 
of four. Among our scaling observables only $\wt{m}_{\rm PS}L$, $\eta_{\rm PS}L$ 
and $\wt{\eta}_{\rm PS}L$ depend on $\ca$, and the dependence vanishes as 
$\alpha \to \pi/2$. The situation is hence similar to the case of $\cv$.    
Concerning the error associated with the use of one-loop values for $\cT$, 
we just remark that the non-perturbative estimate given in Ref.~\cite{gupta} 
at $\beta=6$ is not much larger than the one-loop value, which is about 0.02,
and affected by large relative uncertainties.

In view of these remarks and for the sake of simplicity, in our 
analysis we have neglected all uncertainties on the values
of the improvement coefficients.

\item {\bf Uncertainties on $L\mr$} \\
By extra simulations at the same bare parameters as those of the point A in 
Table~\ref{tab:sim1}, but with values of $\kappa$ such that $L\mr \simeq 0.010$
and $L\mr \simeq 0.034$, we estimate the derivatives of our meson observables
with respect to $L\mr$. All derivatives are of order 1 and compatible with
zero within errors. These estimates are employed to (slightly) move the 
central values of the observables, so that the nominal renormalization
condition $L\mr = 0.020$ is exactly matched, and to add quadratically 
to the statistical errors the uncertainties arising from the 
quoted error on $L\mr$. The effect of this correction is however very tiny,
as well as the modification of the errors on the meson observables.  

\item {\bf Uncertainties on $L\mur$} \\
By extra simulations at the same bare parameters as those of the point A in
Table~\ref{tab:sim1}, but with values of $\muq$ such that $L\mur \simeq 0.140$
and $L\mur \simeq 0.168$, we estimate the derivatives of our meson observables
with respect to $L\mur$. All derivatives take values between 0.5 and 3, with
relative errors less than 10\%. The estimated uncertainties are employed to add 
quadratically to the statistical errors the uncertainties arising from the
quoted error on $L\mur$. The corresponding increase of the statistical errors
on the meson observables is significant only in the case of $\eta_{\rm{PS}}$
and $\widetilde{\eta}_{\rm{PS}}$.

\item {\bf Uncertainties on $L/r_0$} \\
By an extra simulation at the same bare parameters as those of the point A in
Table~\ref{tab:sim1}, but for $\beta=6.06$ and a value of $\kappa$ such
to maintain $L\mr \ll L\mur$, we finally estimate the derivatives of our meson 
observables with respect to $L/r_0$. We obtain estimates of order 1 for the 
would-be meson masses, of about 0.3 for $\eta_{\rm{PS}}$ and 
$\widetilde{\eta}_{\rm{PS}}$ and smaller than 0.1 for $\eta_{\rm{V}}$ and
$\widetilde{\eta}_{\rm{V}}$. The relative errors on these estimates are of about 
10\%, except for the derivatives of $\eta_{\rm{V}}$ and $\widetilde{\eta}_{\rm{V}}$
which have much larger relative errors. Also in this case, the estimated derivatives 
(or conservative upper bounds on them) are employed to add quadratically to the 
statistical errors the uncertainties arising from the quoted error on $L/r_0$. The
corresponding increase of the statistical errors on the meson observables is
typically not larger than half the standard deviations in Table~\ref{tab:jo2}. 

\end{itemize}

\begin{table}[htb]
\vspace{0.25cm}
\begin{center}
\begin{tabular}{ccccc}
\hline
set &  $m_{\rm PS}L$ & $m_{\rm V}L$ & $\eta_{\rm PS}L$ & $\eta_{\rm V}$ \\
\hline
 A   & 1.861(18)      & 2.688(22)      & 0.5434(53)          & 0.1534(29)
\\
 B   & 1.805(22)      & 2.652(29)      & 0.5570(68)          & 0.1565(38)
\\
 C   & 1.831(22)      & 2.646(30)      & 0.5514(67)          & 0.1615(40)
\\
 D   & 1.828(29)      & 2.652(37)      & 0.5499(82)          & 0.1596(50)
\\
\hline
set  &  $\widetilde{m}_{\rm PS}L$ & $\widetilde{m}_{\rm V}L$
& $\widetilde{\eta}_{\rm PS}L$ & $\widetilde{\eta}_{\rm V}$        \\
\hline
 A   & 1.704(12)      & 2.395(18)      & 0.5686(47)          & 0.1823(30)
\\
 B   & 1.667(13)      & 2.337(22)      & 0.5793(53)          & 0.1889(39)
\\
 C   & 1.680(13)      & 2.345(22)      & 0.5751(52)          & 0.1933(40)
\\
 D   & 1.665(16)      & 2.324(29)      & 0.5756(62)          & 0.1944(51)
\\
\hline
\end{tabular}
\caption{{\sl
Final results for our meson observables: the standard deviations given in 
parentheses account for all the uncertainties that we discuss in the text.
}}\label{tab:jo3}
\end{center}
\end{table}

\begin{figure}[htb]
\begin{center}
\epsfig{file=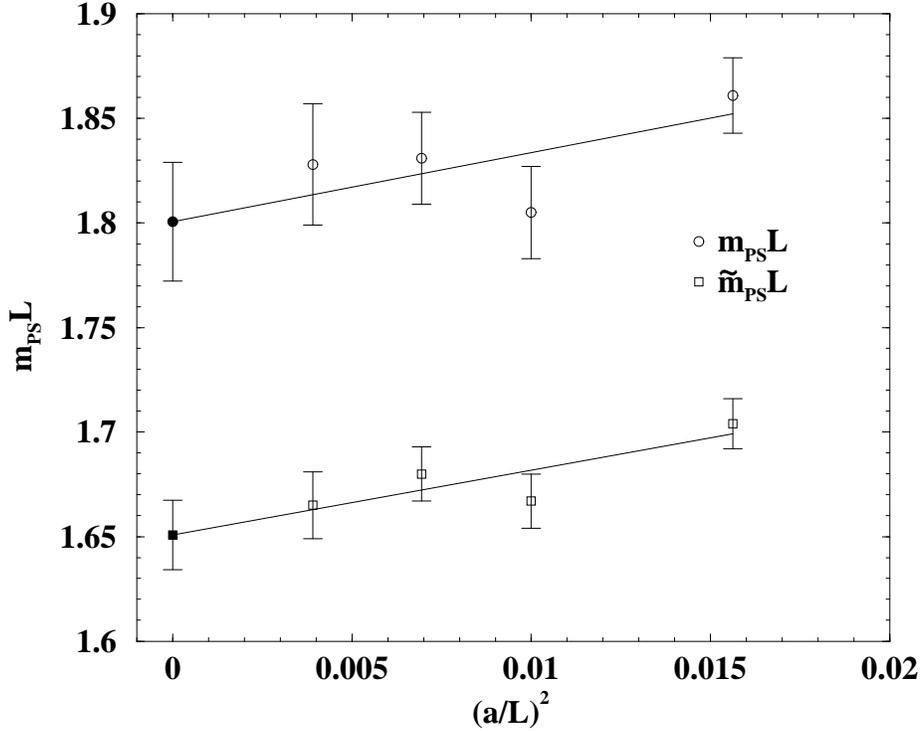,width=10.0cm,angle=-90}
\caption{{\sl
Scaling behaviour and continuum extrapolation for 
$m_{\rm PS}L$ and $\widetilde{m}_{\rm PS}L$.
}}\label{fig:mps}
\end{center}
\end{figure}

\begin{figure}[htb]
\begin{center}
\epsfig{file=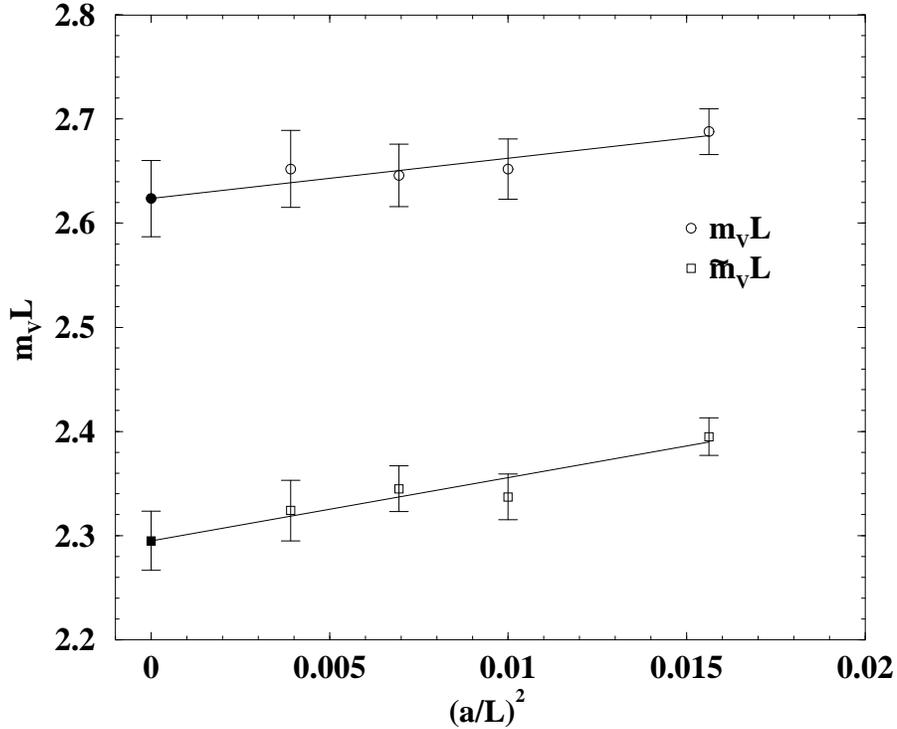,width=10.0cm,angle=-90}
\caption{{\sl
Scaling behaviour and continuum extrapolation for 
$m_{\rm V}L$ and $\widetilde{m}_{\rm V}L$.
}}\label{fig:mV}
\end{center}
\end{figure}

\begin{figure}[htb]
\begin{center}
\epsfig{file=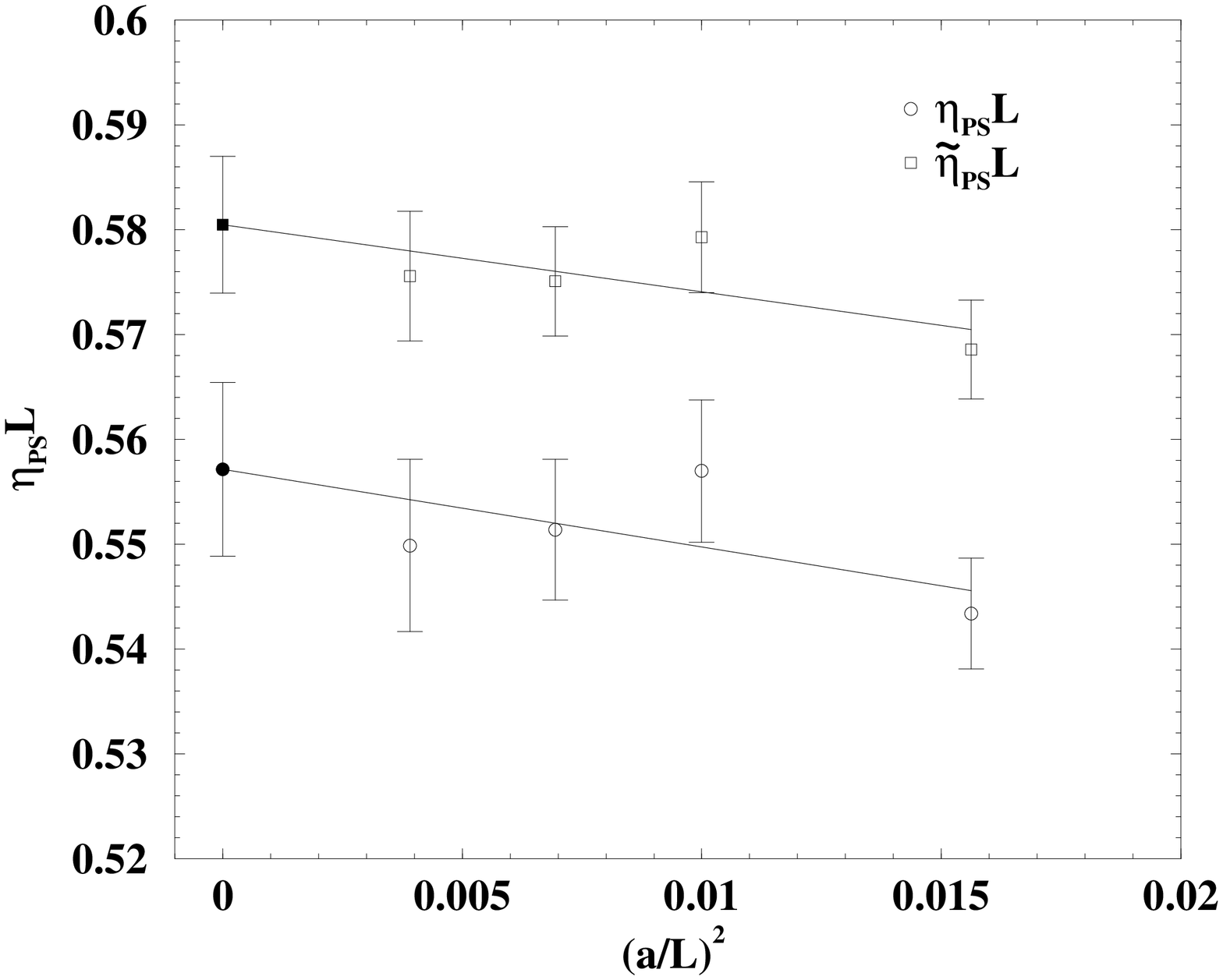,width=10.0cm,angle=-90}
\caption{{\sl
Scaling behaviour and continuum extrapolation for 
$\eta_{\rm PS}L$ and $\widetilde{\eta}_{\rm PS}L$.
}}\label{fig:etaps}
\end{center}
\end{figure}

\begin{figure}[htb]
\begin{center}
\epsfig{file=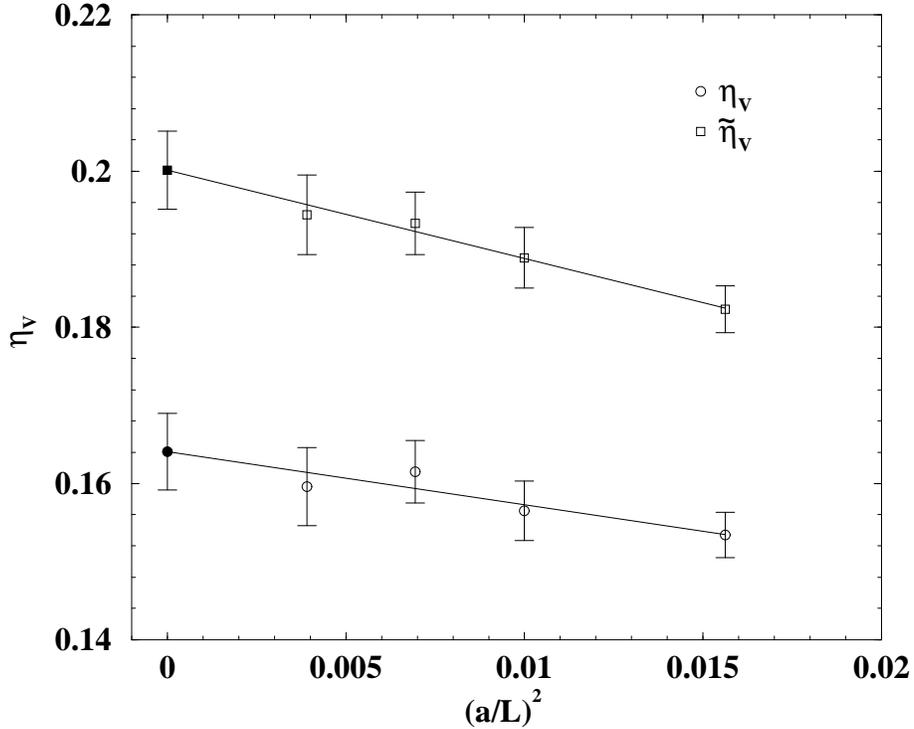,width=10.0cm,angle=-90}
\caption{{\sl
Scaling behaviour and continuum extrapolation for 
$\eta_{\rm V}$ and $\widetilde{\eta}_{\rm V}$.
}}\label{fig:etaV}
\end{center}
\end{figure}

We present the outcome of this analysis in Table~\ref{tab:jo3} and plot
the same results versus $(a/L)^2$ in the Figures~\ref{fig:mps}--\ref{fig:etaV}.
In view of the almost complete implementation of non-perturbative O($a$) 
improvement, for  each observable we perform a least squares fit of the
data to a polynomial of first order in $(a/L)^2$. The fit line and
the extrapolated continuum values with their uncertainty are also shown
in the mentioned plots. The values of the $\chi^2$ per degree
of freedom are of order 1 for all the fits we did, and in most cases smaller
than 1. The observed scaling violations of our meson observables are hence
apparently compatible with an O($a$) improved approach to the continuum limit.

In Table~\ref{tab:dev} we quote the extrapolated continuum limit values
for all our meson observables, together with the relative deviations of 
these values from the values measured at $a\simeq 0.1~{\rm fm}$ (point A of 
Table~\ref{tab:sim1}). These relative deviations, which can be considered 
as a measure of the size of the lattice artifacts, are indeed fairly small
and of the same order of magnitude as the analogous relative deviations
observed under similar conditions at $\alpha = 0$ \cite{Jo}. We remark
that the largest relative scaling violation is observed for $\wt{\eta}_{\rm V}$,
an observable which was not considered in Ref.~\cite{Jo} and might be
affected by residual O($a$) cutoff effects due to the poor knowledge of
$\cT$. Indeed, both $\wt{\eta}_{\rm V}$ and $\wt{m}_{\rm V}L$, which are
the only observables that depend on $\cT$, show relative scaling violations
larger than their $\cT$-independent counterparts, $\eta_{\rm V}$ and $m_{\rm V}L$.

%

%
\begin{table}[htb]
\vspace{0.25cm}
\begin{center}
\begin{tabular}{cccc}
\hline
  $m_{\rm PS}L$ & $m_{\rm V}L$ & $\eta_{\rm PS}L$ & $\eta_{\rm V}$ \\
\hline
  1.801(28)       & 2.624(37)      & 0.5572(83)[15]      & 0.1641(49)[6]
\\
  3.3\%           & 2.4\%          & 2.5\%               & 6.5\%
\\
\hline
  $\widetilde{m}_{\rm PS}L$ & $\widetilde{m}_{\rm V}L$
& $\widetilde{\eta}_{\rm PS}L$ & $\widetilde{\eta}_{\rm V}$        \\
\hline
  1.651(17)       & 2.295(28)      & 0.5805(65)[15]      & 0.2001(50)[7]
\\
  3.2\%           & 4.4\%          & 2.1\%               & 8.9\%
\\
\hline
\end{tabular}
\caption{{\sl
Continuum limit values of our meson observables and their relative 
deviations from the values at $\beta=6$. The additional errors due to small systematic
uncertainties in the non-perturbative estimates of $\zv$ and $\za$ 
are shown in square brackets.  
}}\label{tab:dev}
\end{center}
\end{table}
%





\subsection{Residual O($a$) cutoff effects}

The fact that the dependence on $a/L$ of our scaling observables,
for the considered values of $\beta$ and within relative statistical 
errors of $1$--$2 \%$, can consistently be
described as purely quadratic does not imply the complete
absence of cutoff effects linear in $a/L$. We can only conclude that they are
small enough to be not clearly visible in our data, which in some cases might 
also be due to accidental cancellations between different O($a$) cutoff effects.
These remarks do not apply to the case of $r_{\rm PCVC}$, where the cutoff effects 
linear in $a/L$ are expected to be fully negligible, and are actually not seen
in the data with statistical errors of a few permille.

On the remaining scaling observables, the only quantitatively significant O($a$) 
effects may arise from possibly inappropriate values given either to $\ct$ and $\ctildet$
or to $\tbv$ and $\tba$. In the latter case the effects are proportional to $a\muq$,
which is never larger than 0.01, whereas the effects arising from the use of 
perturbative values for $\ct$ and $\ctildet$ are suppressed in the limit of large $T$ 
(and are thereby irrelevant for physical applications).
The lacking knowledge of the non-perturbative values of these improvement coefficients
makes any estimate of these uncancelled O($a$) effects rather subjective. However, 
in order to disentangle the various residual O($a$) effects and get a rough idea of 
their magnitude, we have looked at the influence of independent variations of $\tbv$,
$\tba$, $\ct$ and $\ctildet$ on our observables.
  
When varying $\tbv$ from its one-loop value, eq.~(\ref{1l_tildeb}), to a value of
order 1, we observe at $\beta=6$ a change of a few standard deviations in 
$\widetilde{m}_{\rm {PS}}$,
$\eta_{\rm{PS}}$ and $\widetilde{\eta}_{\rm{PS}}$, and no changes elsewhere. 
If we instead vary $\tba$ from its one-loop value, eq.~(\ref{1l_tildeb}), to a
value of order 1, we note at $\beta=6$ a change of about one standard deviation 
in $L\mr$,
$\eta_{\rm{V}}$ and $\widetilde{\eta}_{\rm{V}}$, and negligible changes in all
other quantities.  

Varying the values of $\ct$ and $\ctildet$ requires to perform extra simulations.
We have hence repeated the simulations at the bare parameters of point A and point C 
of Table~\ref{tab:sim1} by changing either the value of $\ct$ or the value of $\ctildet$.
Following Ref.~\cite{Jo},  
we have chosen the new values of $\ct$ and $\ctildet$ so that $\ct-1$ and $\ctildet-1$  
are about 2 and 10 times, respectively, larger than the values employed
in our scaling test. A variation of $\ct$ by this amount induces no statistically
significant changes in our meson observables. Under the mentioned variation of
$\ctildet$, 
\[ \ctildet = 1-0.018g_0^2 \quad \longrightarrow \quad \ctildet = 1-0.180g_0^2 \, ,  \]
we do see statistically significant changes in a few observables, namely 
$\widetilde{m}_{\rm{PS}} L$, $\eta_{\rm{PS}}L$ and $\widetilde{\eta}_{\rm{PS}}L$.     
At $\beta=6$ (point A of Table~\ref{tab:sim1}) the observed changes amount to 
$4$--$ 6 \%$ of the mean values of these observables, whereas at $\beta=6.26$
(point C of Table~\ref{tab:sim1}) the relative changes are smaller by a factor 
of 1.5, i.e. they scale proportionally to $a/L$ as expected. 
If one propagates the changes observed at $\beta=6$ and $\beta=6.26$ to the
other values of $\beta$ considered in this work (by assuming that they are
proportional to $a/L$), and then tries to fit the resulting values of 
$\widetilde{m}_{\rm{PS}} L$, $\eta_{\rm{PS}}L$ and $\widetilde{\eta}_{\rm{PS}}L$
for $\ctildet = 1-0.180g_0^2 $ 
to a constant plus a term $\propto (a/L)^2$, still reasonably good fits
are obtained.  

These quantitative checks about the influence of reasonable changes of $\tbv$,
$\tba$, $\ct$ and $\ctildet$ on the meson scaling observables, 
together with the possibility of accidental 
cancellations, show that our results in Subsection 4.3 are compatible with
the presence of residual O($a$) cutoff effects, which may individually have
a relative magnitude of a few percents. 
On the other end, fitting the results of Table~\ref{tab:jo3} to a
polynomial of second order in $a/L$ yields continuum limit values 
that are consistent with those in Table~\ref{tab:dev},  but with much
larger uncertainties. As it is clear from Figures~\ref{fig:mps}--\ref{fig:etaV},    
the coefficient of the term $\propto a/L$ always takes values that
are consistent with zero within errors.

\section{Conclusions}
We have presented a scaling test for some representative hadronic observables 
in the pseudoscalar and vector meson channels computed in quenched lattice twisted mass
QCD with Schr\"odinger functional boundary conditions. We have also studied an
observable that allows to quantify the lattice cutoff effects in the PCVC relation,
which is non-trivial due to $\muq \neq 0$.
To get accurate results at moderate computational effort, an 
intermediate-volume system of physical size $0.75^3\times1.5 \; {\rm fm}^4$ 
has been considered.

We find that in the parameter region specified by $\beta \geq 6$, 
$L\mu_{\rm R}=0.153 \gg Lm_{\rmR}=0.020$ and $T=2L\simeq1.5 \; {\rm fm}$ the 
O($a$) improvement programme of tmQCD can be successfully implemented,
although non-perturbative estimates of $\tbv$ and $\tba$ are still missing.
The size of the observed scaling violations, which range from $0.5 \%$ to
$9 \%$ depending on the observables, is acceptably small and comparable
to the size of the cutoff effects observed in standard lattice QCD with
Wilson quarks.
Studies of lattice tmQCD in larger spatial volume, $L= 1.5$--$ 2.2 \; ~{\rm fm}$, confirm
that the pseudoscalar and vector meson channels can be studied well 
in the chiral regime by simulations at $\beta \geq 6$ with small scaling
violations \cite{paper4}. 

Since this work represents the first non-perturbative study of lattice tmQCD,
it is worthwhile to emphasize that we employ completely standard lattice
techniques and find the computational effort to be essentially the same as 
for Wilson quarks with standard mass parameterization. These findings, which 
have little to do with our particular choice of boundary conditions, are 
basically due to the simple flavour structure of the considered correlation
functions (see Subsection~2.2) and the good performance of the BiCGStab 
solver in the explored parameter region (see Table~\ref{tab:sim1}). 
Following Ref.~\cite{paper2}, the renormalization of the twisted mass 
parameter turns out to be easy in practice, because it can be traced back to the 
renormalization of the non-singlet pseudoscalar density in the massless theory.

Of course it would be interesting to perform similar scaling tests of lattice
tmQCD for different choices of the physical parameters: e.g.~at $L=0.75$ fm
but $T > 2L$, in order to render the O($a$) improvement of the SF boundaries
unimportant, or at higher values of $\mur r_0$, as well as for the lattice theory
with two twisted light quarks and one heavier standard quark, which is
relevant for the study of kaon physics \cite{B_K_proj,paper1}. Some work in 
this direction is planned for the near future.  

\subsection*{Acknowledgements}
This work is part of the ALPHA Collaboration research programme. 
We thank DESY and INFN for allocating computer time on the APE100 machines 
to this project, and the staff of the DESY computer centre at Zeuthen for 
their support.
Moreover we are grateful to R.~Sommer for several discussions.

M.~Della~Morte and R.~Frezzotti thank MURST for support.

%

%
%
\end{document}